# Power loss and temperature distribution in coil of PFC inductor with air gap for multimode operation

Rafał Kasikowski

Lodz University of Technology, Faculty of Electrical, Electronic, Computer and Control Engineering, Institute of Electronics, al. Politechniki 10, 93-590 Łódź

**Abstract:** Power converters inherently display non-linear load characteristics, resulting in a high level of mains harmonics, and hence the necessity of implementing Power Factor Correction (PFC). Active PFC circuitry typically comprises an inductor and a power switch to control and alter the input current so that it matches, in shape and phase, the input voltage. This modelling of the waveforms can be performed by means of distinct conduction modes of the PFC inductor. The digital controller implemented in the constructed and investigated boost-type PFC converter can be programmed to operate in discontinuous conduction mode (DCM), continuous conduction mode (CCM), or a combination of the two. The individual modes of operation, via distinct PFC inductor current waveforms, impact the overall efficiency of power conversion and, by extension, temperature distribution in the magnetic component. This paper investigates how the examined conduction modes bear on distinct power-loss mechanisms present in the PFC inductor, including high-frequency eddy-current-generating phenomena, and the fringing effect in particular. As demonstrated herein, the DCM operation, for the set output power level, exhibits exacerbated power dissipation in the winding of the inductor due to the somewhat increased RSM value of the current and the intensified fringing magnetic flux at an air gap. The latter assertion will undergo further, more quantitatively focused research. Finally, the construction of the coil was optimised to reduce power loss by diminishing eddy-current mechanisms.

## 1. Introduction

Power Factor Correction (PFC) modules are an intrinsic part of all medium-to-high-power AC/DC converters. This is due to the international regulation standards concerning current harmonics, which essentially require power supplies falling into a certain classification and with a power consumption greater than stated limits to be equipped with PFC controllers [1]. PFC units effectively alter the current flowing into the converter to match, in shape and phase, the input voltage. The finely tuned input waveforms ensure the high power factor necessary to maximise the active power drawn from the mains and provide very low Total Harmonic Distortion (THD), and hence strongly reduced input current harmonics. This can be achieved by active PFC circuitry, which itself can be accomplished by a number of distinct topologies, with a step-up converter being a popular choice [2] (Fig. 1).

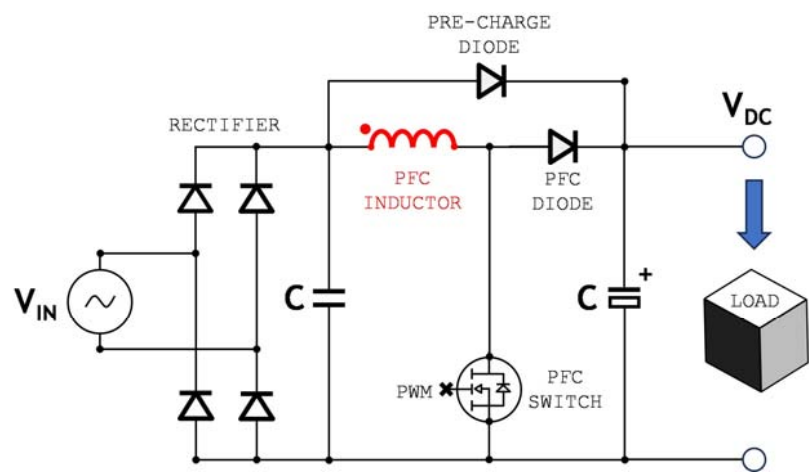

**Fig. 1.** PFC boost topology.
Rys. 1. Architektura przetwornicy PFC typu boost.

The boost-circuit-based PFC requires a relatively low component count and typically operates in discontinuous conduction mode (DCM), continuous conduction mode (CCM), and boundary or critical conduction mode (CrCM), which is frequently combined with the quasi-resonant (QR) operation of a power switch, also referred to as valley switching. The DCM and CrCM/QR modes of operation are primarily incorporated in relatively low-power applications up to 400 W peak output power, chiefly due to the high peak currents present in the inductor and switch. The CCM mode is principally utilised in high-power applications [3]. The individual modes of operation are illustrated in Fig. 2, whereas the quasi-resonant switching is shown in Fig. 3. In the case of PFC inductor current waveforms for individual modes, each is pictured over an identical interval covering a single half-cycle of the sinusoidal input voltage. Fig. 3 illustrates a small number of switch-on-switch-off cycles for the PFC power switch. The design of the PFC choke, the frequency of operation, and the chosen conduction mode determine the number of individual on-off cycles of the switch within a single half cycle of the AC input voltage.

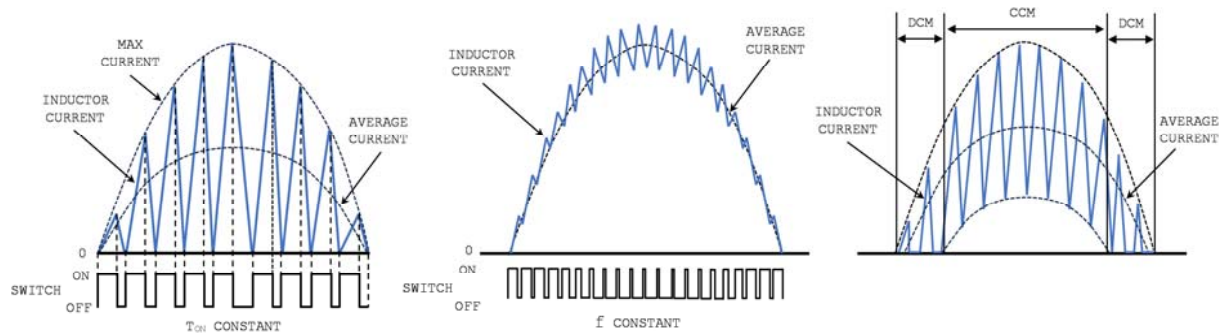

**Fig. 2.** PFC inductor current waveforms for different operating modes. CrCM (left), CCM (middle), mixed-mode operation (right).
Rys. 2. Przebiegi prądu cewki PFC dla różnych trybów pracy. CrCM (po lewo), CCM (po środku), tryb mieszany (po prawo).

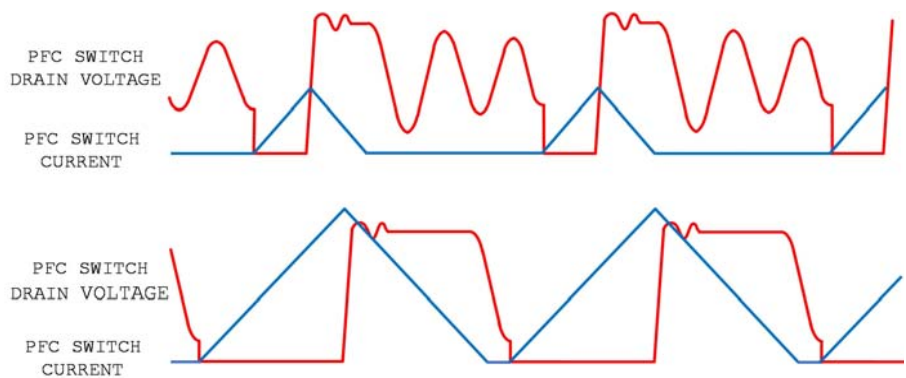

**Fig. 3.** PFC switch drain voltage (red) and PFC inductor current (blue) waveforms for quasi-resonant (QR) operation. Light loading (top), heavy loading (bottom).
Rys. 3. Przebiegi napięcia na drenie tranzystora PFC (na czerwono) i prądu cewki PFC (na niebiesko) dla quasi-rezonansowego trybu pracy. Dla lekkiego obciążenia (góra), dla dużego obciążenia (dół).

The QR mode is implemented to boost the efficiency of power conversion by keeping switch-on losses at a possible minimum level. The PFC switch is turned on at the valley of the ringing of the drain voltage after the magnetic component is already demagnetized (PFC inductor current is zero). The valley at which the switch-on occurs depends on the level of the power converted and the parameters of the PFC inductor [4]. The design of PFC inductors for individual conduction modes is somewhat complex and generally involves several factors to be taken into consideration, such as a desired range of switching frequencies for each operation, magnetic flux density swing, or thermal performance. The latter, along with efficiency of power conversion, are the key criteria behind the successful design of the magnetic component. Typically, a PFC inductor comprises a ferrite core that is suitably gapped to extend the power handling capacity of the component and prevent the core from reaching saturation and shedding its magnetic properties. The minimization of power dissipated in the windings of a PFC inductor is rather problematic due to the complex nature of the current waveforms and insidious eddy-current-inducing mechanisms, including the fringing effect. The growing demand for reduced size and greater power density frequently forces designers, in order to procure more compact and efficient designs, to opt for magnetic cores of reduced dimensions in relation to the power they convert, hence the necessity of implementing relatively large air gaps. This introduces extra power loss into the design since the size of an air gap is directly linked to the extent to which the fringing magnetic flux at the air gap loops out of the intended magnetic path and interacts with the windings [5], [6], [7]. As pictured in Fig. 2 and Fig. 3, individual modes of operation are characterised by distinct PFC inductor current waveforms, and therefore the impact of a given mode on power loss in windings and, by extension, temperature distribution can be observed. These different current waveforms correlate to specific evolutions of magnetic flux in the core and thus may have noticeably distinct impacts on fringing-effect power loss. Infrared (IR) thermography allows for the identification of power-loss sources and the qualitative ascertainment of power loss [8], [9].

## 2. Constructed PFC module

To investigate power loss and temperature distribution in the windings of a PFC inductor with an air gap for multimode operation, the PFC/LLC converter featured in Fig. 4 was designed and constructed. The selection of the PFC controller was determined by the functional requirement of facilitating the conduction modes outlined in the previous section. The design featured the TEA2017 CCM/DCM/QR PFC/LLC digital controller IC [10], where the PFC module of the IC can be configured to operate in three different modes: DCM/CrCM/QR, CCM fixed frequency, or mixed-mode operation [11]. The LLC (resonant topology) module of the controller, although implemented, was disabled and had no impact on the functionality of the PFC stage or the measurements carried out. To register the impact on the way in which temperature and, by extension, power loss are spread out across the winding for each of the examined conduction modes, it was ensured that the constructed PFC stage and the magnetic component incorporated in it operated at constant input and output parameters. The converter was connected to a 150-VAC power source, and the output voltage of the PFC stage was set to 400 V. The PFC inductor itself comprised a 65-turn double-layer coil of 0.7mm diameter, mounted on an E42/21/15 core of 3F3 ferrite material [12], and centre-gapped to 2.1 mm. This particular geometry of the magnetic component resulted in an inductance of 650 µH.

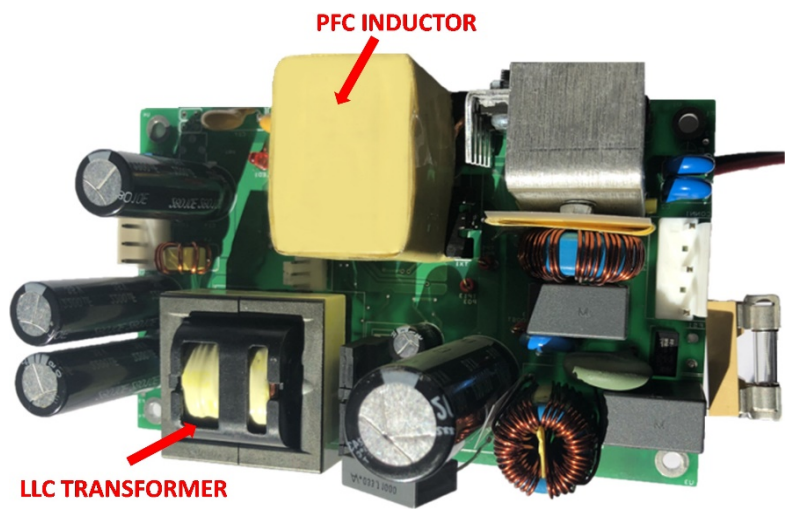

**Fig. 4.** Constructed converter with PFC module.
Rys. 4. Skonstruowana przetwornica z modułem PFC.

In the process of the measurements, the inductor was subjected to two different modes of operation: the DCM/CrCM/QR and the mixed-mode operation (DCM/CrCM/QR/CCM), as pictured in Fig. 2. The oscilloscope screenshots presenting the actual PFC switch drain voltage and inductor current waveforms captured at the nominal power level are shown collectively in Fig. 5. The screenshots present a number of switch-on-switch-off cycles of the PFC power switch registered at the peak of the input voltage sinusoid. The RMS values of the inductor current over the entire cycle of the input voltage for each of the conduction modes were measured with a RIGOL MSO5074 digital oscilloscope. The figures registered by the device were about 2.31 A and 2.69 A for the mixed mode and the DCM/CrCM/QR, respectively. One can immediately surmise that the continuous conduction mode for the set output power level of 300 W should have somewhat reduced ohmic power loss resulting solely from the DC resistance of the wire used in the construction of the magnetic component, as this mechanism of power dissipation is directly proportional to the square of the RMS value of the inductor current. Apart from the opposition to the flow of electric current in the form of DC resistance, power loss occurring in the winding is brought about by eddy-current-inducing phenomena, namely the skin effect [13], proximity effects [14], and the fringing effect. These three eddy-current mechanisms have their effects exacerbated at high frequencies and significantly complicate the design of PFC inductors. The impact is particularly evident in magnetic components comprising many turns and layers as undesirable heating is generated. As the investigated PFC inductor featured a single-strand and multi-layer coil, a considerable share of the total power loss in the wires was expected to be caused by the so-called AC resistance associated with the described phenomena. Consequently, at the final stage of the research, the 0.7-mm-diameter wire, initially employed in the construction of the coil, was substituted with a multi-strand litz-type wire [15], where a carefully arranged pattern causes the current in the coil to be distributed homogenously throughout the strands so that the net current in each is effectively the same.

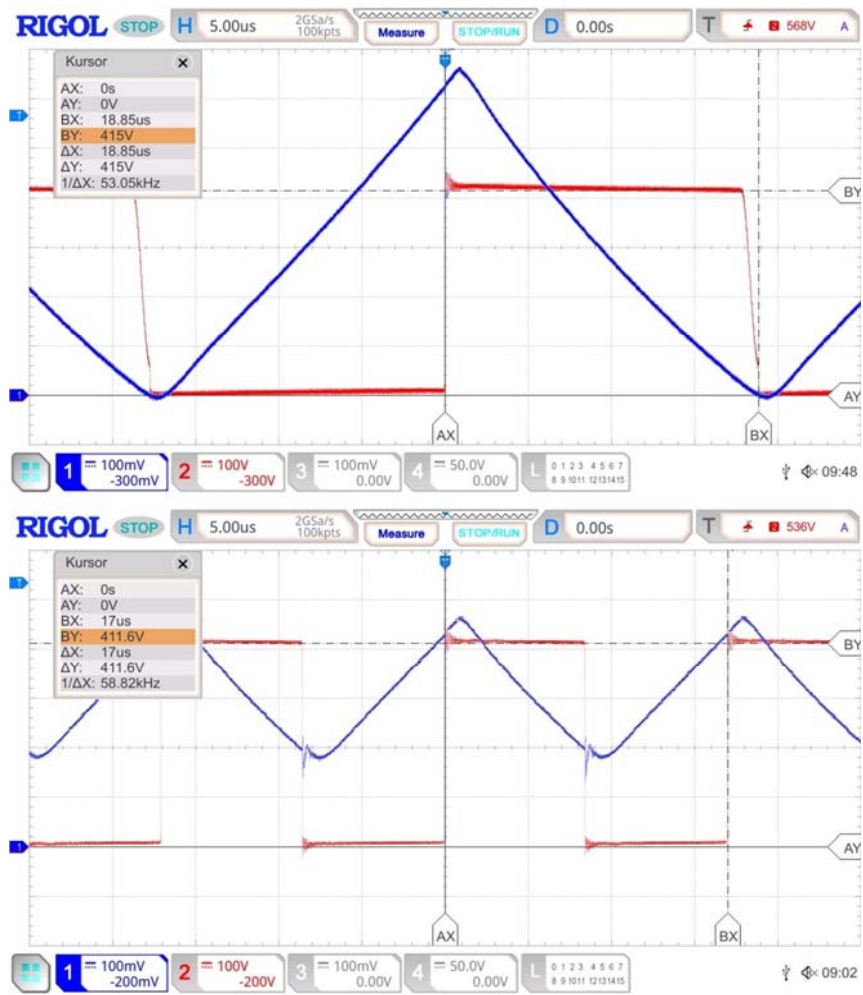

**Fig. 5.** PFC switch drain voltage (red) and inductor current (blue) for DCM/CrCM/QR (top) and DCM/CrCM/QR/CCM (bottom).
Rys. 5. Przebiegi napięcia na drenie tranzystora PFC (na czerwono) i prądu cewki PFC (na niebiesko) dla trybu pracy DCM/CrCM/QR (góra) oraz DCM/CrCM/QR/CCM (dół).

### 3. Thermal measurement setup

The thermal measurement setup arranged for the purpose of temperature distribution and power loss evaluation is shown in Fig. 6. Prior to the measurements, the magnetic device was mounted away from any heat-radiating components in the converter so as to eliminate their thermal impact and assure the accuracy of the measurements. The PFC module was continuously run at 300 W for each of the investigated operating modes until steady-state temperatures in the system were reached. At this point, a series of thermographs were registered by a MWIR Cedip Titanium cooled photon camera operating at a 25-Hz frame rate. The thermal images extracted from the sequence filmed show the temperature distribution in the inductor right before the measurements were brought to a close. The ambient of 27.0 °C remained unchanged throughout the procedure. The thermograms registered for DCM/CrCM/QR and DCM/CrCM/QR/CCM (mixed-mode) operations are presented in Fig. 7 and Fig. 8, respectively. The thermographs were juxtaposed with each other to clearly show that the operation comprising the CCM mode (Fig. 8) displays lower overall temperatures and hence a reduced power loss. This is mainly due to the reduced RMS value of inductor current for the mixed-mode. One can also notice that the ends of the coil show the lowest temperature, while the maximum temperature is registered in the area directly above the air gap. This distinct pattern can be clearly attributed to the fringing effect phenomenon at the air gap [16].

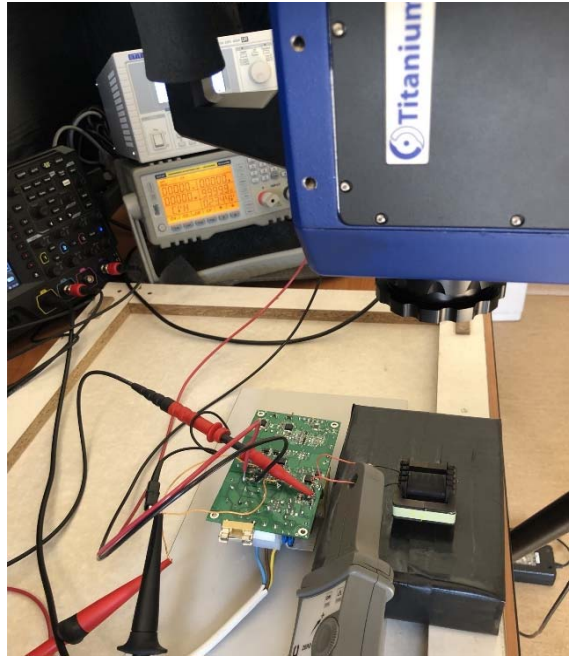

**Fig. 6.** Thermal measurements setup.
Rys. 6. Stanowisko do pomiarów termicznych.

The evolution of temperature along the middle column, particularly the extended range of temperature change between the ends of the coil and the area directly above the air gap, indicates that the fringing effect might be somewhat exacerbated for the DCM/CrCM/QR (Fig. 7), but this claim requires further investigation. As can be read from the curves, the increase in temperature along the coil is about 36.0 °C and about 29.5 °C for the DCM/CrCM/QR and the mixed mode, respectively. This signifies that the temperature range was boosted by about 6.5 °C for the former, a finding that is rather unlikely to be entirely accounted for by the slightly increased RMS value of the current. The RMS value can be held, at least partly, responsible for the upsurge of about 12.0 °C in the minimal temperature registered for the DCM/CrCM/QR configuration, as the DC resistance has, essentially, a nearly homogenous distribution along the wire.

The actual value of the DC resistance of the coil was determined by forcing a DC current of 1 A through the wire, measuring voltage between the ends of the coils, and applying Ohm's law. This allows for the estimation of power dissipated in the winding solely due to the RSM value of the inductor current for each of the conduction modes:

For the DCM/CrCM/QR, Eq. 1:

$$P_{DC_{DCM/QR}} = R_{DC} \cdot I^2_{DCM/QR} = 0.211\,\Omega\,\cdot 2.69^2 A = 1.52\,W \tag{1}$$

For the DCM/CrCM/QR/CCM, Eq. 2:

$$P_{DC_{DCM/QR/CCM}} = R_{DC} \cdot I^2_{DCM/QR/CCM} = 0.211\,\Omega\,\cdot 2.31^2 A = 1.31\,W \tag{2}$$

Furthermore, the frequency of the waveforms, one of the key variables in eddy-current power loss mechanisms, for each of the modes differs significantly. At the moment of entering the CCM, the frequency of the mixed-mode operation remains relatively constant, whereas the frequency at which the inductor current rises and falls for the DCM/CrCM/QR gradually decreases with increasing output power. As indicated by the AX/BX markers in Fig. 5, a full cycle of the PFC power switch for the latter operation lasts, for the output power of 300W, approximately twice as long as the corresponding cycle for the mixed mode, about 37.7 μs and about 17 μs, respectively.

Power loss in the material of the core is not frequency independent and will, almost certainly, differ for the investigated conduction modes. The core loss density can be expressed in the form of the extended version of the Steinmetz empirical formula, Eq. 3 [17].

$$P_V = k_f \cdot f^x \cdot \Delta B^y \cdot (ct - ct_1 \cdot T + ct_2 \cdot T^2) \tag{3}$$

where:
$P_v$ is the core loss density, kW/m³
$ct, ct_1, ct_2$ are the ferrite material's coefficients
$T$ is the operating temperature of the core material, C°
$k_f, x, y$ are the ferrite material's coefficients
$f$ is the operating frequency of a given magnetic component, Hz
$\Delta B$ is the peak AC magnetic flux density, T

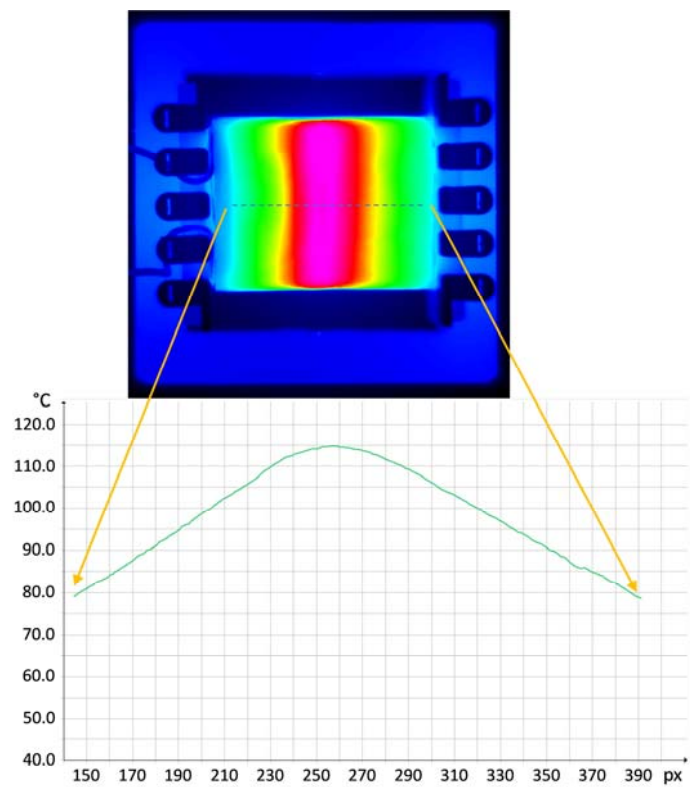

**Fig. 7.** Thermal image of PFC inductor with air gap (top) and temperature distribution along middle column (bottom) for DCM/CrCM/QR.
Rys. 7. Obraz termowizyjny cewki PFC ze szczeliną powietrzną (góra) oraz rozkład temperatury wzdłuż środkowej kolumny (dół) dla trybu pracy DCM/CrCM/QR.

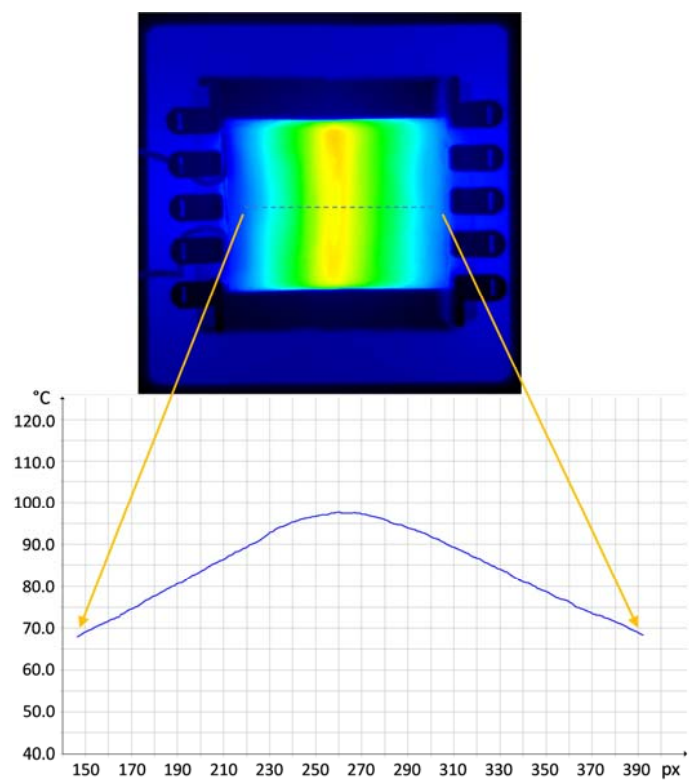

**Fig. 8.** Thermal image of PFC inductor with air gap (top) and temperature distribution along middle column (bottom) for DCM/CrCM/QR/CCM.
Rys. 8. Obraz termowizyjny cewki PFC ze szczeliną powietrzną (góra) oraz rozkład temperatury wzdłuż środkowej kolumny (dół) dla trybu pracy DCM/CrCM/QR/CCM.

The extended Steinmetz empirical formula reveals that the power dissipated in the core can be represented as a function of three variables, namely the equivalent sinusoidal excitation frequency $f$, magnetic flux density $ΔB$, and the operating temperature of the core material $T$. The latter is fairly comparable for the examined modes of operation, as can be inferred from the thermographs (Fig. 7 and Fig. 8), and should not impact the core loss to any significant extent. The magnetic flux density swing, on the other hand, is considerably enlarged for the DCM/CrCM/QR, as the range of changes in this variable follows the variations in the inductor current (Fig. 5). This is counteracted, to a certain degree, by the increased frequency of operation for the mixed mode, and hence it is believed that the power loss in the magnetic material had a rather limited impact on the temperature distribution in the winding; however, a more detailed estimate of core losses for each of the examined cases should be the subject of a further study.

In view of the above, it can be concluded that the shown evolutions of temperature along the middle column for both investigated modes are mainly driven by eddy-current mechanisms, and the fringing effect in particular. To corroborate this claim, the same coil was mounted on an ungapped EMS-0432115-060 core formed out of Sendust magnetic metal powder [18]. Composite materials, such as Sendust, do not have a discrete air gap, but the air gap in them is distributed throughout the entire material. The ungapped core was selected according to its magnetic permeability so that the coil utilised in the experiment yielded exactly the same inductance as its gapped counterpart. Once more, the PFC module was continuously run at 300 W for each of the modes of operation until the thermal steady-state was reached. The temperature distribution along the winding was registered at this point (Fig. 9).

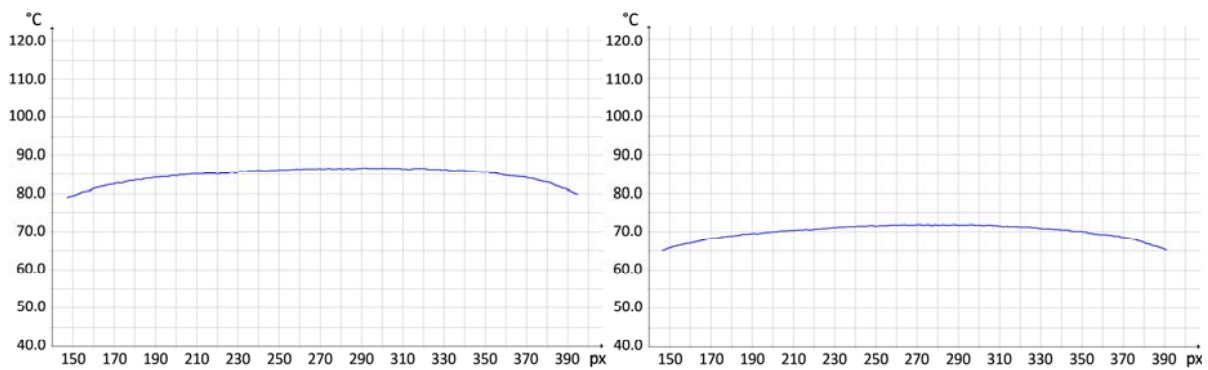

**Fig. 9.** Temperature distribution measured along coil mounted on ungapped core. DCM/CrCM/QR (left), DCM/CrCM/QR/CCM (right).
Rys. 9. Rozkład temperatury wzdłuż uzwojenia umieszczonego na rdzeniu o szczelinie rozproszonej. Tryb DCM/CrCM/QR (po lewo), DCM/CrCM/QR/CCM (po prawo).

As can be noticed, the variation in the temperature measured along the coil for both the DCM/CrCM/QR and the mixed-mode configuration is very much alike, with marginal discrepancies between the two. This is in stark contrast to the thermographs and temperature distributions registered for the gapped core (Fig. 7 and Fig. 8), which feature distinctly elevated temperatures in the vicinity of the air gap. The evolution of temperature between the ends of the winding for the distributed-gap-core inductor is primarily governed by the power loss due to the RMS value of the current and the eddy-current-inducing phenomena, excluding the fringing effect. For that reason, the conclusion that the increase in the temperature range observed between the end of the coil and the area located directly above the air gap is a result of the fringing-effect phenomenon at the air gap should be considered valid, and furthermore, this effect is somewhat exacerbated for the DCM/CrCM/QR conduction mode (Fig. 7).

As the last stage of the research, the optimisation of the winding of the PFC choke was carried out. The 0.7-mm-diameter wire, initially employed in the construction of the coil, was substituted with a litz-type wire. The selection of the wire, its diameter, and the number of strands were based on the method outlined in [19], modified for the purpose of this research. In the presented approach, the frequency of operation, the number of turns, and the geometry of the coil former are needed to compute the optimal parameters of a litz wire. As a consequence, a 0.05-mm, 300-strand litz-type wire was incorporated into the design of the PFC inductor. The steady-state thermograph for the gapped inductor operating in the DCM/CrCM/QR/CCM is presented in Fig. 10. As is clearly visible, the fringing effect and other eddy-current-inducing mechanisms were greatly diminished.

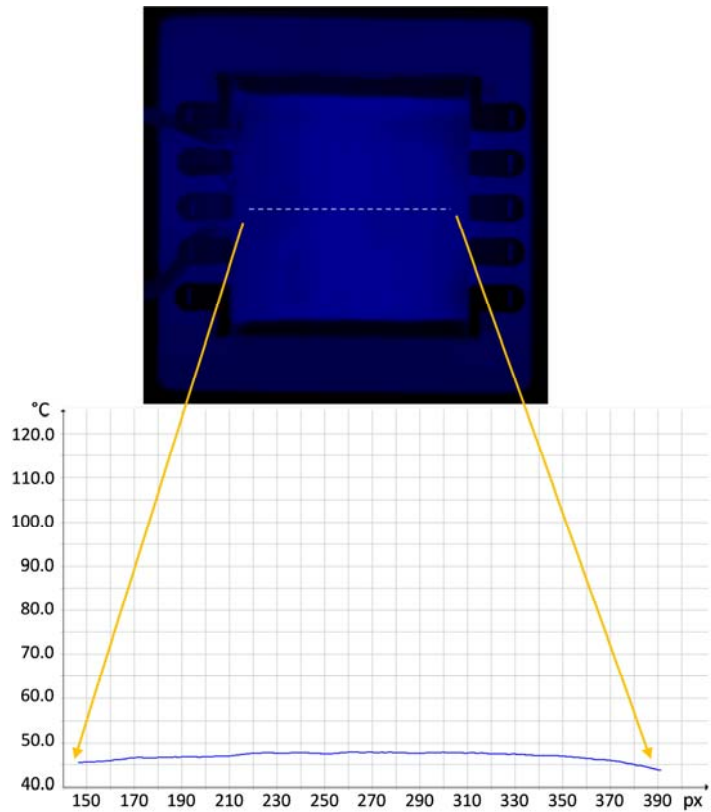

**Fig. 10.** Thermal image of optimized PFC inductor with air gap (top) and temperature distribution along middle column (bottom) for DCM/CrCM/QR/CCM.

Rys. 10. Obraz termowizyjny zoptymalizowanej cewki PFC ze szczeliną powietrzną (góra) oraz rozkład temperatury wzdłuż środkowej kolumny (dół) dla trybu pracy DCM/CrCM/QR/CCM.

## 4. Conclusion

This research work explores the issue of power dissipation and temperature distribution in the winding of a PFC inductor for multimode operation, with a particular focus on magnetic components with an air gap. The digital controller incorporated into the constructed PFC module was configured to operate in two distinct conduction modes: DCM/CrCM/QR and DCM/CrCM/QR/CCM. Subsequently, IR thermographic measurements were performed on the constructed PFC inductor with an air gap. As demonstrated in the course of the research, the inductor, subjected to the investigated operation modes, displayed divergent temperature patterns along the winding, and hence different power loss densities in individual sections of the component. The contribution of distinct power loss sources to the total power loss in the examined magnetic component was discussed. This was followed by an analogous analysis of power loss and thermal performance of the same coil mounted on an ungapped magnetic core. In view of the research results shown, it was concluded that the winding of the PFC inductor operating in the DCM/CrCM/QR mode displayed exacerbated power dissipation due to the fringing-effect phenomenon. This claim will be addressed and receive proper quantitative investigation in further work. Finally, the initially investigated coil was optimised to reduce power loss originating primarily from high-frequency eddy-current-inducing mechanisms.

**Funding**

This research was funded by Narodowe Centrum Nauki as part of the MINIATURA 6 project, grant number - 562623, under registration number - 2022/06/X/ST7/01370.

# Straty mocy i rozkład temperatury w uzwojeniu cewki PFC ze szczeliną powietrzną dla różnych trybów pracy.

**Streszczenie:** Przetwornice impulsowe z natury charakteryzują się nieliniową charakterystyką obciążenia, co skutkuje wysokim poziomem harmonicznych prądu wejściowego, a co za tym idzie koniecznością korekcji współczynnika mocy (ang. Power Factor Correction – PFC). Aktywne układy PFC mają zazwyczaj w swojej budowie element indukcyjny (cewka PFC) oraz klucz elektroniczny w formie tranzystora, które kontrolują i modyfikują prąd wejściowy, tak aby odpowiadał on kształtem i fazą napięciu wejściowemu. Modelowanie przebiegu prądu zwykle jest realizowane za pomocą kilku odmiennych trybów pracy (przewodzenia) cewki PFC.

Cyfrowy układ elektroniczny zaimplementowany w skonstruowanej i następnie badanej przetwornicy PFC podnoszącej napięcie (ang. Boost/Step-up converter) jest programowalny do pracy w trybie nieciągłego prądu (ang. Discountinuous Conduction Mode – DCM), ciągłego prądu (ang. Countinuous Conduction Mode – CCM) lub w ich połączeniu. Poszczególne tryby pracy, a dokładnie różne przebiegi prądu cewki PFC, wpływają na sprawność przetwornicy PFC, a co za tym idzie, na straty mocy i rozkład temperatury w elemencie indukcyjnym. W artykule przedstawiono, w jaki sposób badane tryby pracy wpływają na różne mechanizmy rozpraszania mocy występujące w uzwojeniu cewki PFC, w tym na, występujące dla relatywnie wysokich częstotliwości, zjawiska generujące prądy wirowe, a w szczególności na zjawisko strumienia rozproszenia przy szczelinie powietrznej. Jak zademonstrowano, tryb pracy typu DCM, dla danego obciążenia przetwornicy, wykazuje wyższe straty mocy w uzwojeniu cewki PFC ze względu na powiększoną wartość skuteczną prądu (ang. Root-Mean-Square – RMS) i wzmożony oddziaływanie zjawiska strumień magnetycznego rozproszenia przy szczelinie powietrznej. Ostatnia teza będzie przedmiotem dalszych, ukierunkowanych ilościowo działań badawczych. W ostatnim etapie badań zoptymalizowano konstrukcję uzwojenia cewki, aby zmniejszyć straty mocy poprzez zredukowanie mechanizmów generowania prądów wirowych.

**Słowa kluczowe:** elektronika mocy, przetwornice AC/DC, poprawa współczynnika mocy, cewki, dławiki, termografia, straty mocy

**Rafał Kasikowski, PhD, Eng**

rafal.kasikowski@p.lodz.pl

**ORCID: 0000-0002-2815-1746**

He received the M.Sc. degree in electrical engineering in 2002 from the Technical University of Czestochowa, Poland, and he has also earned the M.Sc. degree in energy engineering in 2013 from the University of East Anglia, Norwich, United Kingdom. In 2021 he received the PhD degree in electronics from Lodz University of Technology, Poland, where he is currently working as a postdoctoral researcher.
His research interests include modelling and optimization of magnetic components in Switch Mode Power Supplies.

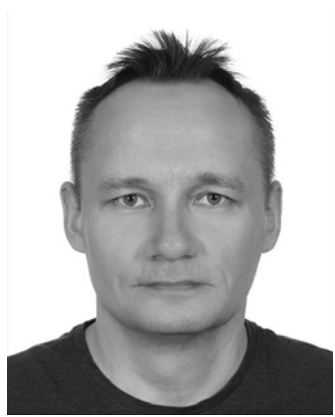